\documentclass[
 twocolumn,floatfix,showpacs,preprintnumbers,
 amsmath,amssymb,pre]{revtex4}

\usepackage{graphicx}
\usepackage{bm}

\renewcommand{\v}[1]{{\boldsymbol #1}}

\DeclareGraphicsExtensions{.eps}
\graphicspath{{figures/}}

\begin{document}

\setlength{\voffset}{3\baselineskip}

\title{Thermodynamic Entropy and Chaos in a Discrete Hydrodynamical System}

\author{Franco Bagnoli}
\email{franco.bagnoli@unifi.it}
\affiliation{Dipartimento di Energetica, Universit\`a di Firenze,\\
 Via S. Marta 3, I--50139 Firenze, Italy\\ also CSDC and INFN, sez.\ Firenze}
\author{Ra\'ul Rechtman}
\email{rrs@cie.unam.mx}
\affiliation{Centro de Investigaci\'on en Energ\'\i a, Universidad Nacional
 Aut\'onoma\\
 de M\'exico, Apdo.\ Postal 34, 62580 Temixco, Mor., Mexico}

\date{\today}

\begin{abstract}
We show that the thermodynamic entropy density is proportional to the largest 
Lyapunov exponent
(LLE) of a  discrete hydrodynamical system, a  deterministic two-dimensional 
lattice gas
automaton. The definition of the LLE for cellular automata is based on the
concept of  Boolean derivatives and is formally equivalent to that of
continuous dynamical systems. This relation is justified 
using a Markovian model. In an irreversible process with an initial density 
difference
between both halves of the system, we find that Boltzmann's $H$ function is
linearly related to the  
expansion factor of the LLE, although the latter is more sensitive 
to the presence of traveling waves. 
\end{abstract}


\maketitle


\section{Introduction}

The relation between thermodynamics and the underlying chaotic properties
of a system is of great relevance in the foundations of statistical
mechanics~\cite{lebowitz73,zaslavsky04} and has attracted much interest. What is the 
relation between chaotic dynamics and thermodynamics? Is chaos relevant for 
thermodynamics? Some interesting results have been found for some models. In the
case of a family of simple liquids, including Lennard--Jones, a relation has been 
found between the Kolmogorov--Sinai entropy and the thermodynamic 
entropy~\cite{dzugutov98}. Lorentz gases have been extensively studied and
relationships between chaotic dynamical properties and transport coefficients have been 
established~\cite{gaspard90,evans90,gaspard95,dorfman97,dorfman02}. The discussion 
of this problem for other simple models and maps is 
extensive~\cite{gaspard97,latora99,tasaki99,gilbert00,tasaki00}. 

Lorentz gases are essentially one-particle systems.
In this paper we present a simple
model of a gas with a large number of interacting particles
and find interesting relations
between dynamic and thermodynamic quantities, somewhat motivated by 
Bernoulli systems. For these, the Kolmogorov-Sinai entropy is on the one hand the sum of the
positive Lyapunov exponents, and on the other, the thermodynamic entropy 
scaled by a time constant corresponding to the correlation 
time~\cite{billingsley82,cornfeld82}. 

The model we study is a deterministic
lattice gas cellular automaton (LGCA). LGCA are simple models 
with hydrodynamical behavior~\cite{hardy73,frisch86,frisch87}. In particular, the 
$D2Q9$ LGCA is a two dimensional model with nine velocities and is one of the simplest 
models where equilibrium thermodynamics can be found with a non trivial 
temperature~\cite{bagnoli93}. For cellular automata, Lyapunov exponents can be 
defined using Boolean derivatives~\cite{bagnoli92}. This definition is formally similar
to that of continuous dynamical systems, a long time average of the logarithm
of the linearized expansion factor~\cite{ott93}. We also look for a relation between 
Boltzmann's $H$ function in a simple irreversible process, and the logarithm of the 
expansion factor.

The paper is organized as follows. In Sec.~\ref{sec:d2q9} we present a deterministic
version of the $D2Q9$ model together with an introduction to Lyapunov exponents
of cellular automata.
In Sec.~\ref{sec:thermo} we discuss the equilibrium thermodynamics of the $D2Q9$ models
and in Sec.~\ref{sec:results} we show there is a close relationship between the 
equilibrium entropy of the model and its largest Lyapunov exponent. This relationship
is explained by finding the Kolmogorov-Sinai entropy of a Markov chain which relates
the thermodynamic entropy to the largest Lyapunov exponents. We also show
that Boltzmann's $H$ function goes to its equilibrium value in an irreversible 
process in the same way the Lyapunov exponent does. We end with a discussion
on why these quantities are related.


\section{The $D2Q9$ model} 
\label{sec:d2q9}

The $D2Q9$ model is defined on a two dimensional square
lattice~\cite{chen89}. The evolution is in discrete time steps where unit mass particles at
every site $\v{r}$ can move with one of nine velocities $\v{c}_0=(0,0)$,
$\v{c}_1=(1,0)$,  $\v{c}_2=(0,1)$, $\v{c}_3=(-1,0)$,
$\v{c}_4=(0,-1)$, $\v{c}_5=(1,1)$, $\v{c}_6=(-1,1)$,
$\v{c}_7=(-1,-1)$, $\v{c}_8=(1,-1)$.
The state of the automaton is given by the set
of occupation numbers $\v{s}(t) = \{s_k(\v{r},t)\}$, where
$s_k(\v{r},t)=1$ (0) indicates the presence (absence) of a particle with velocity $\v{c}_k$ 
at site $\v{r}$ and time $t$. An exclusion principle forbids the presence of more than
one particle in a given site and a given time with a given velocity. 

The time evolution of the
system is given by collision and streaming operations. In the collision step,
the particles at every site change their velocities conserving mass, momentum and energy. 
In the streaming operation particles move to neighboring sites according to their velocities.
\begin{figure}
 \includegraphics[width=.9\columnwidth]{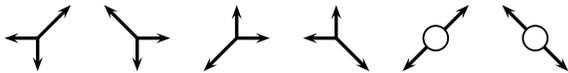}
 \caption{\label{fig:collision} All these states have the same number
  of particles,  momentum and energy. An arrow represents
  the presence of a particle with the velocity of the arrow, the open circle a 
  particle at rest.}
\end{figure}

Since the number of states for a given site is finite ($2^9$), the local collision 
operator $C$ is generally implemented as a look-up table. Given a local configuration, 
the conservation constraints may not define the outgoing state completely as the example 
in Fig.~\ref{fig:collision} shows. If any one of the six states shown is chosen 
as the input state, any one of the other six states can be the output state. Therefore, 
the $C$ look-up table has several columns for all the possible output states. In order 
to make the automaton deterministic, we first find for all input states, the number of 
output states. We then find the least common multiplier of the number of output states 
and construct a new $C$ look-up table with this number of columns. These are filled
by repeating the output states the required number of times. The least common 
multiplier is sixty, so the rows which have any one of the input states of 
Fig.~\ref{fig:collision} are repeated ten times. At the beginning we assign  
an integer random number $\eta(\v{r})$ to each site, to be used in choosing 
the column from which the output state is chosen (quenched disorder). That is
\begin{equation}
 s_k(\v{r},t+1)=C_k(s_0(\v{r},t),\dots,s_8(\v{r},t),\eta(\v{r})).
\end{equation}
The choice of the quenched disorder is analogous to the random disposition of 
scatterers in a wind-tree or other similar Lorentz gases~\cite{ehrenfest11,ernst89}.

\begin{figure}
 \includegraphics[width=8cm]{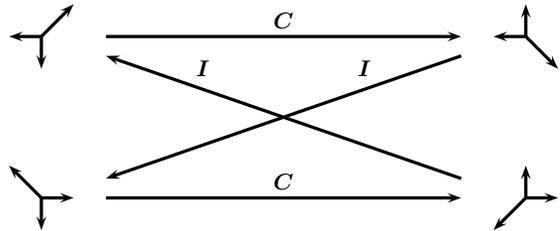}
 \caption{\label{fig:reversecoll} In this example, the input state at
 some fixed site is the one shown on the left of
 Fig.~\protect{\ref{fig:collision}} and in the column number assigned
 to this site, the output state is the fourth one of that Fig. These
 are shown in the upper part of this Fig.  Then, by inverting the
 velocities we get the state on the lower left and the collision table
 should contain in the same column for this input state the output
 state shown on the lower right that upon inversion of the velocities
 yields the original incoming state.} 
\end{figure}

The evolution of the model can also be made reversible. In order to do so, the collision 
table must satisfy the condition that in every column 
$s=\v{I}\v{C}\v{I}\v{C}(s)$ holds where $\v{I}$ denotes the operator
that inverts the velocities of any state $s$. This means that a
collision $\v{C}$ followed by an inversion $\v{I}$ and another
collision and inversion leaves the state unchanged when it is taken
from the same column in the collision table as we show in
Fig.~\ref{fig:reversecoll}.  Once we have a look-up table that
is deterministic as described above, the elements in each row are rearranged to
satisfy the reversibility condition. 

In order to introduce the concept of Lyapunov exponents for such a
system, it is convenient to simplify the  notation used. We use the
index $n$ to indicate the position $\v{r}$ and the velocity $k$, with
$n=1,\dots,9L$, where $L$ is the number of sites of the automaton.
A configuration of the LGCA at a given time is given by $9L$
occupation numbers (bits), and its evolution can be seen as an
application of a set of Boolean functions
\begin{equation}{\label{eq:fn}}
 s_n(t+1)=F_n(\v{s}(t)).
\end{equation}
The functions $F_n$ represent different entries of
the collision table, and differ in the velocity index $k$ and
quenched disorder $\eta(\v{r})$. However,
since the distribution of the disorder is uniform, and, as shown in
the following, the correlations among variables decay very fast, the
system is translationally invariant at a mesoscopic level.

Let $\v{s}(0)$ and $\v{x}(0)$ be two initially close configurations,
for example all the components of $\v{x}(0)$ may be equal to those
of $\v{s}(0)$ except for one.  We define the bitwise difference
between these two configurations with the term ``damage''. The
smallest possible damage is one and the damage vector
$\v{v}(0)$ has a one in the component where $\v{s}(0)$ and
$\v{x}(0)$ differ and has zeroes in all the others. 
If this damage
grows on average during time, the trajectory is unstable with
respect to the smallest perturbation. However, due to the discrete
nature of LGCA, defects may annihilate during time evolution,
altering the measure of instability of trajectories. The correct way
of testing for instability is that of considering all possible ways
of inserting the smallest damage in a configuration, using as many
replicas as the number of components of the configuration, and
letting them evolve for one time step counting if the number of
damages has grown or diminished. The ensemble of all possible
replicas with one damage each is the equivalent of the tangent space of
discrete systems.

The task of computing the evolution in tangent space is clearly
daunting, but by exploiting the concept of Boolean
derivatives~\cite{bagnoli92,bagnoli92a} it is possible to develop a formula
very similar to the one used in continuous systems. The Boolean
derivative is defined by
\begin{equation}\label{eq:boolean}
 \dfrac{\partial F_n(\v{s})}{\partial s_p}=%
  \left|F_n(\dots,s_p,\dots)-F_n(\dots,1-s_p\dots)\right|
\end{equation}
with $n,p=1,\dots,9L$. This quantity measures the sensitivity
of the function $F_n$ with respect to a change in $s_p$. The
Jacobian matrix $J(\v{s})$ has components $J_{np}=\partial
F_n(\v{s})/\partial s_p$.

We now  consider the map
\begin{equation}\label{eq:t-map}
 \v{v}(t+1)=J(\v{s}(t))\v{v}(t)
\end{equation}
with $\v{v}(0)$ as mentioned above.
It is easy to check that $|\v{v}(t)|=\sum_i v_i(t)$ is the number of
different paths along which a damage may grow in tangent space
during time evolution, {\em i.e.}, with the prescription of just one
defect per replica~\cite{bagnoli92}. If there is sensitivity with
respect to initial conditions, one expects that 
$|\v{v}(T)|/|\v{v}(0)|\sim \exp(\lambda_{_T} T)$ for large $T$ where 
$\lambda_{_T}$ is the largest finite time
Lyapunov exponent (LLE). It then follows that
\begin{equation}\label{eq:lambda}
 \lambda_{_T}=\dfrac{1}{T}\sum_{t=1}^{T-1}\log u(t)%
          =\langle \log u\rangle_{_T}
\end{equation}
with $u(t)=|\v{v}(t)|/|\v{v}(t-1)|$ the expansion factor of the LLE. The definition 
should include the limit when $T\to\infty$ but in practice we always evaluate the
finite time LLE with $T$. The LLE depends in principle on the initial configuration 
$\v{s}(0)$, initial damage $\v{v}(0)$ and quenched disorder $\eta$, but in practice it 
assumes the same value for all trajectories corresponding to the same macroscopic 
observables when $T$ is sufficiently large. The LLE of CA as defined above has been 
used to classify elementary and totalistic Boolean cellular 
automata~\cite{bagnoli92,bagnoli94}.


\section{Thermodynamics of the $D2Q9$ model}
\label{sec:thermo}

The single particle velocity distribution functions $f_k(\v{r},t)$ are defined
as the average number of particles at site $\v{r}$ with velocity 
$\v{c}_k$ at time $t$ over $R$ samples that share the same quenched disorder $\eta$ and
macroscopic constraints with different microscopic initial
configurations. 
 
We checked that the single site two-particle correlation function 
factorizes into the product of single particle distributions before
and after the collision, in equilibrium and out of equilibrium
conditions. In equilibrium, the correlation function decays to zero
for just one lattice spacing. Out of equilibrium, starting with very
different configurations in the two halves of the system, the
correlation function, still being quite small, exhibits a correlation
length of some lattice spacings for a short time. This 
corresponds to the coherent motion of particles in a shock wave,
where the local density is near to zero or nine. However, this
correlation quickly disappears; although the motion is correlated at a
macroscopic level, as soon as the local density of particles is
different from the extremes (for which the collision table has few
output configurations) the velocities quickly decorrelate.

The thermodynamic entropy density can be found analytically in the thermodynamic 
limit as follows~\cite{salcido91}.  Let
\begin{equation}
 \label{eq:constraints}
 N=\sum_k N_k,\qquad E=\sum_k\epsilon_k N_k
\end{equation}
with $N$ and $E$ the number of particles and total energy respectively, $N_k$ the number
of particles in direction $k$, and $\epsilon_0=0$, $\epsilon_{1,2,3,4}=1/2$, 
$\epsilon_{5,6,7,8}=1$. The number density $n$, the equilibrium density functions
$f_k$ and the energy density $e$ are
\begin{equation}
 n=\dfrac{N}{L},\qquad  f_k=\dfrac{N_k}{L},\qquad  e=\dfrac{E}{L}
\end{equation}
with $L$ the number of sites in the lattice. The microcanonical partition function 
$\Omega$ is
\begin{align}
 \Omega=\int\prod_k dN_k&\delta(N-\sum_k N_k)%
           \delta(E-\sum_k\epsilon_k N_k)\times\nonumber\\
       &\omega(N_0,\dots,N_8)
\end{align}
with $\delta$ the Dirac-$\delta$ function and
\begin{equation}
 \omega=\prod_k\dfrac{L!}{N_k!(L-N_k)!}.
\end{equation}
the number of microscopic states in Fermi-Dirac statistics. 

The thermodynamic entropy is $S(E,N,L)=\log\Omega(E,N,L)$. Using Stirling's 
approximation and the 
Fourier transform of the $\delta$ function
\begin{equation}
 \label{eq:S}
 S=\log\left(L^9\int\prod_k df_k\int dp_1dp_2\exp(L\sigma)\right)
\end{equation}
with
\begin{align}
 \sigma=&-\sum_k\left[f_k\log f_k+(1-f_k)\log(1-f_k)\right]\nonumber\\
	&+ip_1\left[n-\sum_k f_k\right]+ip_2\left[e-\sum_k\epsilon_k f_k\right].
\label{eq:sigma}
\end{align}
The integrals of Eq,~(\ref{eq:S})  can be evaluated using the saddle point method 
and the result is exact in the thermodynamic limit. The thermodynamic limit entropy 
density $s$ is
\begin{align}
 s(e,n)&=\lim_{L\to\infty}\dfrac{1}{L}S(eL,nL,L)\nonumber\\
       \label{eq:s}
       &=-\sum_k\left[\hat{f}_k\log\hat{f}_k+(1-\hat{f}_k)\log(1-\hat{f}_k)\right].
\end{align}
In the last expression $\hat{f}_0,\dots\hat{f}_8,\hat{p}_1$, and $\hat{p}_2$ maximize 
$\sigma$ given by Eq.~(\ref{eq:sigma}). 

The quantities $p_1$ and $p_2$ are related to temperature and chemical potential.
Taking the partial derivative of $\sigma$ with respect to $f_k$ and equaling to zero 
we find that
\begin{equation}
 \label{eq:fermi}
 \log\left(\dfrac{\hat{f}_k}{1-\hat{f}_k}\right)=-i\hat{p}_1-i\hat{p}_2\epsilon_k,
\end{equation}
and that
\begin{equation}
 \hat{f}_k=\left[1+\exp(i\hat{p}_1+i\hat{p}_2\epsilon_k)\right]^{-1}.
\end{equation}
Using Eq.~(\ref{eq:fermi}) the entropy density is
\begin{equation}
 s=-\sum_k\left(\log(1-\hat{f}_k\right)+i\hat{p}_1+i\hat{p}_2\epsilon_k.
\end{equation}
The Euler equation is
\begin{equation}
 s=\dfrac{e}{T}+\dfrac{P}{T}-\dfrac{\mu N}{T}
\end{equation}
where $P$ is the pressure and $\mu$ is the chemical potential. Comparing the last 
two expressions
\begin{align}
 \dfrac{1}{T}&=i\hat{p}_2,\\
 \label{eq:mu}
 -\dfrac{\mu}{T}&=i\hat{p}_1,\\
 \dfrac{P}{T}&=-\sum_k\log(1-\hat{f}_k).
\end{align}
Now, the equilibrium distribution functions are
\begin{equation}
 \hat{f}_k=\left[1+\exp(\epsilon_k/T-\mu/T)\right]^{-1}.
\end{equation}

\begin{figure}
\begin{center}
\includegraphics[width=0.9\columnwidth]{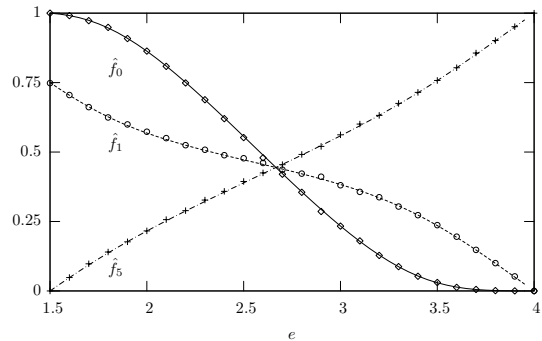}
\end{center}
\caption{\label{fig:fk} Distribution functions in equilibrium as a function of
$e$ for $n=4$. The continuous curves are the calculated values, the symbols are
the results of numerical simulations. For $e=e_F$ there is a rest particle
in all sites and no fast particles moving along the diagonals. The other three
particles can occupy one of the four slow directions. For $e=e_M$ there are no rest
or slow particles, the four particles are moving along the diagonals.}
\end{figure}

The distribution functions $\hat{f}_0$, $\hat{f}_1$, and $\hat{f}_5$ define the
equilibrium state since $\hat{f}_1=\hat{f}_2=\hat{f}_3=\hat{f}_4$ and 
$\hat{f}_5=\hat{f}_6=\hat{f}_7=\hat{f}_8$. These satisfy the conservation equations
of mass end energy and maximize $\sigma$. We also note that these distributions
satisfy another condition equivalent to the latter. From Eq.~(\ref{eq:fermi}) for 
$k=1,5$ and for $k=0$ and Eq.~(\ref{eq:mu})
\[
 \dfrac{\mu}{T}=\log\left(\dfrac{f_0}{1-f_0}\right)=%
       \log\left(\dfrac{f_1}{1-f_1}\right)^2\left(\dfrac{1-f_5}{f_5}\right).
\]
Then
\begin{equation}
 f_1^2(1-f_0-f_5)=f_0f_5(1-2f_1).
\end{equation}
This last expression, together with mass and energy conservation determine
the equilibrium values of the distribution functions. In Fig.~\ref{fig:fk} we show
the values of the equilibrium distribution functions as continuous curves, together
with result of numerical simulations. The agreement between numerical simulations and computed values of the probability distributions is perfect, and this confirms that the system is a perfect gas. Further investigations on correlations in out of equilibrium conditions are reported in the following Section.

\section{Results and discussion}
\label{sec:results}

\begin{figure}
\begin{center}
\includegraphics[width=1\columnwidth]{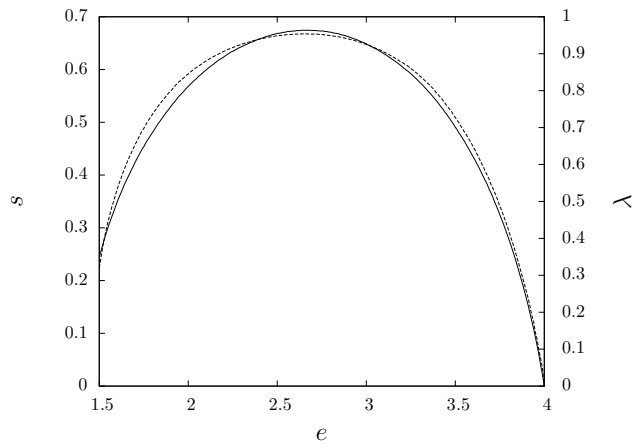}
\end{center}
\caption{\label{H-lambda} Entropy density $s$ (solid line) and LLE
$\lambda$ (dashed line) for $n=4$, simulations with $R=40$ in a
$40\times 40$ lattice.}
\end{figure}

\begin{figure}
\begin{center}
\includegraphics[width=0.9\columnwidth]{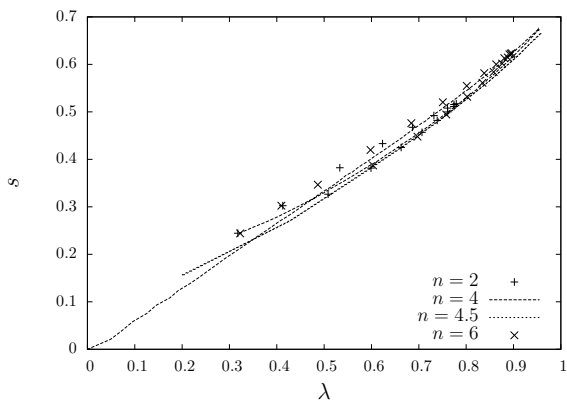}
\end{center}
\caption{\label{fig:H-lyap1} We show $s$ versus $\lambda$ for
$n=2$ (diamonds), $n=4$ (solid line), $n=4.5$ (dotted line) and $n=6$
(pluses). The simulations are performed for different values of $e$
from $e_\text{F}(n)$ to $e_\text{M}(n)$.}
\end{figure}

In Fig.~\ref{H-lambda} we show the entropy density $s=S/L$ and the LLE $\lambda$ as a 
function of $e$ for fixed $n$ both calculated numerically. The entropy
density is found by substituting the numerical values of $f_k$, for example those
of Fig.~\ref{fig:fk}, in 
Eq.~(\ref{eq:s}). The entropy density grows for 
$e_\text{F}\leq e\leq e^\ast$ and then decreases for $e^\ast\leq e\leq e_\text{M}$ with
$e^\ast=2n/3$. The energy density $e^\ast$ is the value for which $f_0=f_1=f_5$.
The largest Lyapunov exponent $\lambda$ shows the same behavior having a maximum also 
at $e^\ast$. The results shown suggest that $s$ is proportional to $\lambda$. This is 
emphasized in Fig.~\ref{fig:H-lyap1} where the data lie near a straight line for
different values of $n$. 
This result shows that the LLE grows with the number of available states
measured by the equilibrium entropy.

The proportionality between the thermodynamic entropy density and the largest Lyapunov
exponent can be understood in the framework of the stochastic approximation of a chaotic 
dynamics~\cite{dorfman99}.
In this approximation, the dynamics is considered at discrete time intervals, and generates a 
discrete and finite partition of coarse-grained states labeled by an index  $i$, with 
$i=1,\dots, M$.
The evolution is represented by a Markov chain where $W_{ij}$ is the transition 
probability from state $j$ to state $i$ and $\sum_i W_{ij}=1$. 
The asymptotic probability distribution is denoted by $p_i$, with $\sum_i p_i=1$ and 
$\sum_j W_{ij} p_j = p_i$. 
The Kolmogorov-Sinai entropy $K$ per time interval $\tau$ is~\cite{dorfman99,ott93}
\begin{equation}
 K = -\sum_j p_j\sum_i W_{ij}\log W_{ij},\label{ks}
\end{equation}
and the entropy density $s$ is 
\begin{equation}
s=-\sum_i p_i \log p_i. 
\end{equation}

For Bernoulli systems the equilibrium is reached in just one time step,
$W_{ij} = p_j$, and trivially $K=\tau^{-1}s$~\cite{gaspard93}. For a somewhat more general process,
we assume that $p_i=1/M$ (microcanonical distribution)
so that $s=\log M$. We further assume that the transition matrix is irreducible
(so that the system is ergodic)
 and that in each row and column there are $\alpha_i M$ non zero equal entries with values 
$1/\alpha_i M$, $0<\alpha_i\leq 1$. Then
\begin{equation}
 K = \langle\log\alpha\rangle+\log M,
\end{equation}
where $\langle\log\alpha\rangle$ is the average of $\log\alpha_i$. 

If we further assume the validity of the Pesin relation~\cite{pesin77},
$K$ equals the sum of positive Lyapunov exponents. For many Hamiltonian systems and symplectic maps, 
the Lyapunov spectrum takes the form $\lambda_i \simeq \lambda_0(1-i/(2N))$, where
$\lambda_i$ is the $i$-th Lyapunov exponent, $i=0,\dots,2N-1$  and the number of particle $N$ is large.
In these cases, the Kolmogorov-Sinai entropy per degree of freedom is proportional to the largest Lyapunov
exponent $\lambda_0$. The proportionality constant, however, may depend on the value of 
control parameters, namely the energy.

The shape of the Lyapunov spectrum is roughly linear for the product
of random matrices with the structure of (weakly) locally coupled Hamiltonian chaotic systems~\cite{paladin86} and for coupled  nonlinear oscillators~\cite{ruffo86}.0
If we assume that also in other cases the shape of the spectrum (which in general is not linear, see for instance Ref.~\cite{beijren04} for the hard spheres case) does not change with the energy, it follows that $s$ and $\lambda$ are linearly related. This last assumption is probably the weakest one, at least for LGCA for which the Lyapunov spectrum is ill-defined, and we think it is the main reason for the discrepancy from linearity in Figure~\ref{fig:H-lyap1}.

\begin{figure}
\begin{center}
\includegraphics[width=1\columnwidth]{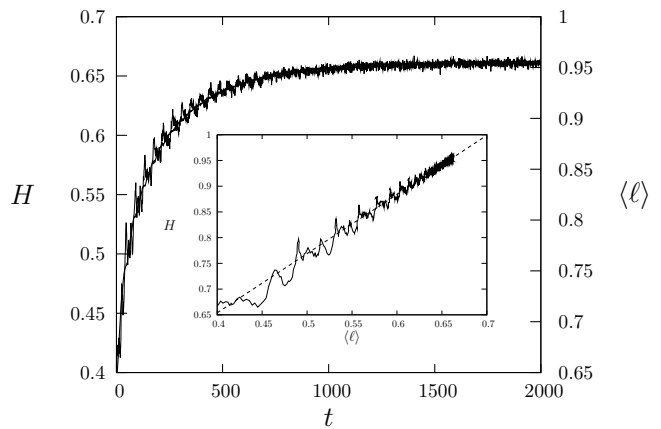}
\end{center}
\caption{\label{H} Boltzmann function $H$ (thick line) and
Lyapunov expansion factor $\langle \ell\rangle$ (oscillating thin
 line) versus time $t$, simulations with 
 $n_L=7.2$, $e_L=4.8$, $n_R=1.8$, $e_R=1.2$, and $R=40$ on a 
 $40\times 40$ lattice. 
 The inset shows that, disregarding the oscillations of $\langle\ell
\rangle$, there is a linear relation between
 these quantities. The dashed line is the best fit $H=1.15\langle\ell
\rangle+0.2$. }
\end{figure}

In our model, 
the value of the LLE is related to the number of ones in the Jacobian
matrix $J$ defined by Eq.~(\ref{eq:boolean}). This Jacobian matrix
contains the linearized effects of the streaming and collision
operators. Streaming corresponds to a scrambling of the components of
the tangent vector $\v{v}(t)$ and therefore does not alter its norm.
This is left to collisions, when more than one output configurations
are possible. The number of equivalent output configurations in the
collision table is small for the extreme values of number and energy
densities, and larger for intermediate values. Similar considerations
apply to the number of equivalent configurations for a given
macroscopic distribution of density and velocities, and constitute the
microscopic origin of the relation between statistical and
dynamical quantities.

We now discuss an irreversible process where a square lattice is initially 
in an equilibrium state with the left and right sides having different number 
and energy densities $n_L$, $n_R$, $e_L$, and $e_R$. The system evolves toward 
equilibrium by means of damped traveling waves. The single site two-particle 
correlation function, although small, exhibits a correlation
length of some lattice spacings for a short time. 

Macroscopically, one observes a coherent motion of particles in a shock wave,
where the local density is near zero or nine. Although the motion is correlated at a
macroscopic level, as soon as the local density of particles is
different from the extremes (for which the collision table has few
output configurations) the velocities quickly decorrelate. 

Boltzmann's $H$ function is defined by
\begin{equation}
 H(t)=-\sum_{\v{r},k}f_k(\v{r},t)\log f_k(\v{r},t).
\end{equation}
In the numerical simulations the distribution functions are averaged over $R$
samples and the average Lyapunov expansion factor is 
$\langle\ell\rangle=(1/R)\sum_{i=1}^R \log u^{(i)}$. As we show in
Fig.~\ref{H}, the two quantities exhibit similar behavior. The Lyapunov expansion factor 
exhibits more marked oscillations, indicating that it is more sensible to the local 
variations in density. The inset of Fig.~\ref{H} shows that, disregarding oscillations,
$\langle\ell\rangle$ is  linearly  related to $H$ for all the relaxation phase. 

We can identify several time scales in this irreversible process. There is a
fundamental time scale, which is fixed to unity. 
There is also a correlation time scale, which is of the
order of the mean free time which depends on the occurrence of non-trivial
collisions and is larger where the density is either small or large.
During shock waves, locally one may have variations of the density and therefore correlations, as already reported,
of the order of some time steps. A third time scale is given by the oscillations induced by the traveling 
shock waves. This is a dynamical, mesoscopic  scale, that depends on the size of the system, and is revealed by 
the oscillations of $\langle\ell\rangle$ shown in Fig.~\ref{H}. The slowest time scale is given by the 
relaxation to equilibrium, shown both by $H$ and by $\langle\ell\rangle$.

The computation of $\langle\ell\rangle$ is performed using a set of
tangent vectors,  Eq.~\eqref{eq:t-map}, and these vectors
constitute a sort of local memory of the past state. In systems with
local variations of density, as in our system in the presence of
traveling waves, statistical quantities like $H$ depend on the
instantaneous state of the system, while dynamical ones like
$\langle\ell\rangle$ depend also on the variations of this state.
This factor may be the origin of the different relation between
statistical and dynamical quantities in equilibrium and during the
relaxation phase.


\section{Final remarks}

The $D2Q9$ reversible LGCA model we have discussed
exhibits hydrodynamical and thermodynamical behavior. For this model, the 
thermodynamic entropy density is proportional to the largest Lyapunov
exponent. Also, in a simple
irreversible process, Boltzmann's $H$ function is proportional to the expansion
factor of the largest Lyapunov exponent. A simple stochastic coarse grained
dynamics can explain the proportionality between $s$ and $\lambda$. We note 
that this result does not depend on 
the $D2Q9$ dynamics and should hold for other, possibly more realistic models.
Finally, the fact that the thermodynamic entropy density and the LLE of the
model are proportional confirms that the definition of the latter is appropriate.


\section*{Acknowledgments}

The authors thank Roberto Livi and Stefano Ruffo for helpful discussions.
Partial economic support from CONACyT project 25116, and
from the
Coordinaci\'on de la Investigaci\'on Cient\'\i fica de la UNAM is acknowledged.


\end{document}